%% file: ms.tex
\documentclass[runningheads]{llncs}
\usepackage{graphicx}
\usepackage{flushend}
\usepackage{balance}
\usepackage{url}
\usepackage{todonotes}
\usepackage{etex}

\begin{document}
\title{Cyber Taxi:\\ A Taxonomy of Interactive Cyber Training\\ and Education Systems}
%\title{Contribution Title\thanks{Supported by organization x.}}
%
\titlerunning{Cyber Taxi}

\author{Marcus Kn\"upfer\inst{1} \and
Tore Bierwirth\inst{1,2} \and
Lars Stiemert\inst{1,2} \and
Matthias Schopp\inst{1,2} \and
Sebastian Seeber\inst{1,2} \and
Daniela P\"ohn\inst{1,2} \and
Peter Hillmann\inst{1} %\orcidID{0000-0003-4346-4510}
}
\authorrunning{M. Kn\"upfer et al.}
\institute{Universit\"at der Bundeswehr M\"unchen, 85577 Neubiberg, Germany
\email{\{marcus.knuepfer, peter.hillmann\}@unibw.de}
%\url{http://www.unibw.de}
\and
Team localos, Munich, Germany\\
\email{\{tore.bierwirth, lars.stiemert, matthias.schopp, sebastian.seeber, daniela.poehn\}@localos.io}
}
\maketitle              % typeset the header of the contribution
\begin{abstract}
%The abstract should briefly summarize the contents of the paper in 150--250 words.
The lack of guided exercises and practical opportunities to learn about cybersecurity in a practical way makes it difficult for security experts to improve their proficiency.
Capture the Flag events and Cyber Ranges are ideal for cybersecurity training. 
Thereby, the participants usually compete in teams against each other, or have to defend themselves in a specific scenario. 
As organizers of yearly events, we present a taxonomy for interactive cyber training and education. 
The proposed taxonomy includes different factors of the technical setup, audience, training environment, and training setup. 
By the comprehensive taxonomy, different aspects of interactive training are considered. 
This can help trainings to improve and to be established successfully.  
The provided taxonomy is extendable and can be used in further application areas as research on new security technologies.
%Beside this, the architecture is usable in further application areas, as research on malware.
%\keywords{First keyword  \and Second keyword \and Another keyword.}
\end{abstract}

\input{textblocks/introduction}
\input{textblocks/requirements}
\input{textblocks/sota}

\section{\uppercase{Taxonomy}}\label{sec:taxonomy}
\begin{figure*}[!h] 
	\centering
	\noindent
%	\captionsetup{justification=centering}
	\includegraphics[width=1.\linewidth]{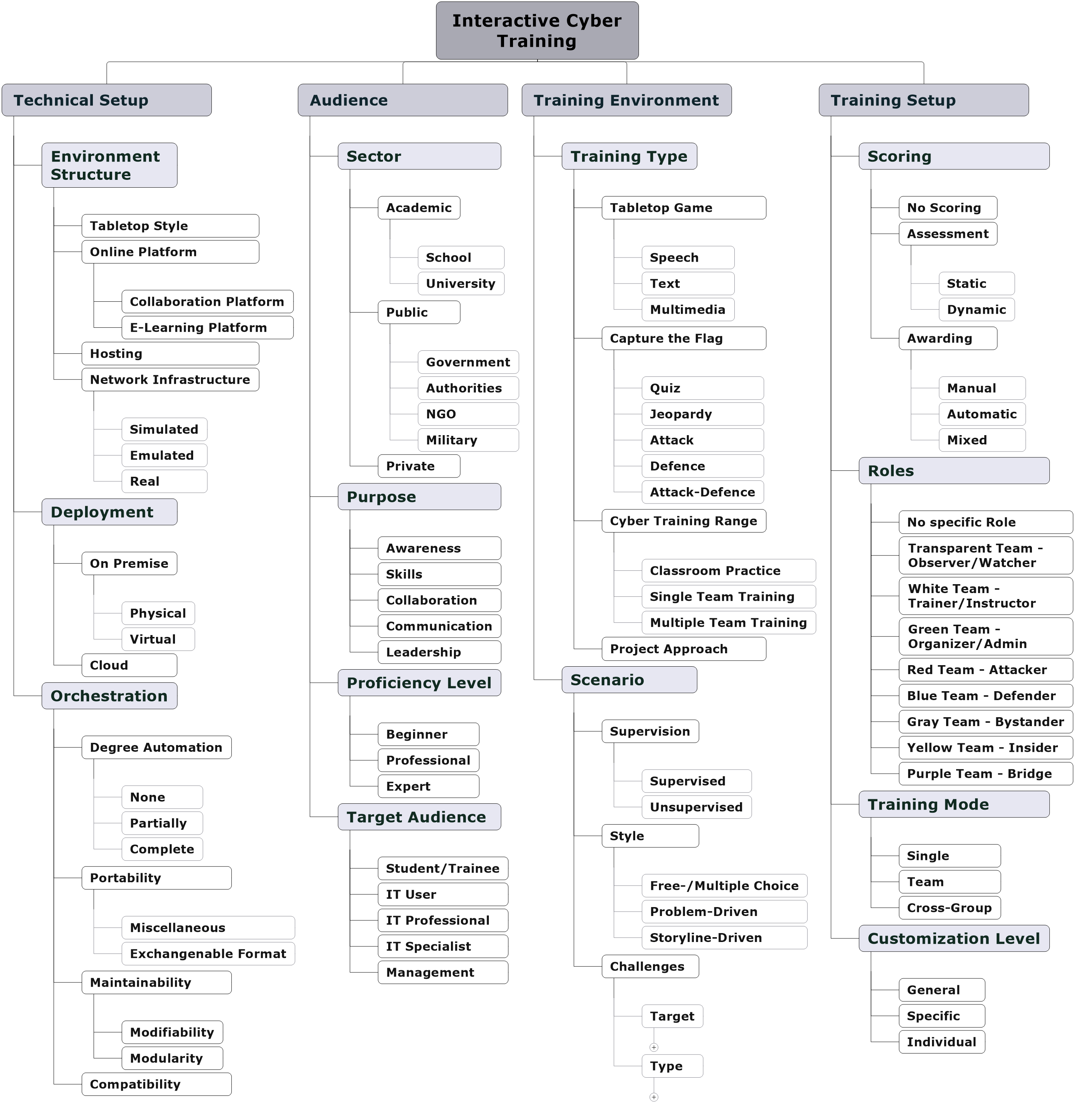}
	\caption{Taxonomy of Interactive Cyber Training and Education Systems}
	\label{fig:taxonomyfull}
\end{figure*}

Within this Section, our taxonomy for ICTE and its components is described in detail.
Figure~\ref{fig:taxonomyfull} provides an overview about the taxonomy. During the design, attention was paid to the complete coverage of all necessary capabilities with regard to the cyber exercise life cycle~\cite{Kick2014} and training competencies~\cite{AcademicExcellence2017}.

\input{textblocks/technical_setup}

\subsection{Audience}
The target of cyber training and education is the audience, which is further characterized in the following. The audience has the characteristics sector, purpose, proficiency level, and target audience.

\subsubsection{Sector:}
The sector from which the audience comes determines the nature of the training. The following categories can be distinguished~\cite{Bishop99whatdo,concordia}.

\begin{itemize}
	\item \textbf{Academic:} This includes universities and schools. The focus is on the principles underlying cybersecurity, ranging from theoretical to applied. 
	\item \textbf{Private:} The private sector and industry focuses more on protecting its investments. The effectiveness of security mechanisms and people are more important than principles they embody.
	\item \textbf{Public:} This includes amongst others Government, NGO, and Military. Cybersecurity is seen as tool to protect the public interest. Hence, it emphasizes on developing policies and systems to implement laws and regulations.
\end{itemize}

\subsubsection{Purpose:}
Purpose answered the question for which reason trainings should be used.
Training can address different objectives which are listed in the following.
\begin{itemize}
	\item \textbf{Awareness:} To raise the awareness in multiple and different security threats. %((2018) Rajamäki - Cybersecurity Education and Training in Hospitals) was für Bedrohungen gibt es und wie kann man Sicherheitsvorfälle vorbeugen
	\item \textbf{Skill:} To recognize the different skill levels of the participants so that can they be improved in a targeted manner.
	\item \textbf{Collaboration:} To improve the cooperation within a team or beyond.
	\item \textbf{Communication:} To increase the efficiency of internal and external communication in case of an incident. % ((2018) Raj - Towards Standards in Undergraduate Cybersecurity Education in 2018) Welche Informationen müssen kommuniziert werden, damit andere Stelle ausreichend informiert sind
	\item \textbf{Leadership:} To improve the management and coordination of the responsible entities.
\end{itemize}

\subsubsection{Proficiency Level:}
Proficiency describes the knowledge of users and what they are able to do.
The proficiency is grouped into three different levels.
%Tore Quelle: https://hr.nih.gov/working-nih/competencies/competencies-proficiency-scale
\begin{itemize}
	\item \textbf{Beginner:} The lowest level. Beginner are limited in abilities and knowledge. They have the possibility to use foundational conceptual and procedural knowledge in a controlled and limited environment. Beginners cannot solve critical tasks and need significant supervision. They are able to perform daily processing tasks. The focus is on learning.
	\item \textbf{Professional:} The mid level. Professionals have deeper knowledge and understanding in specific sectors. For these sectors they are able to complete tasks as requested. Sometimes supervision is needed but usually they perform independently.  
	The focus is on enhancing and applying existing knowledge.
	\item \textbf{Expert:} The highest level. Experts have deeper knowledge and understanding in different sectors. They complete tasks self-dependent and have the possibilities to achieve goals in the most effective and efficient way. Experts have comprehensive understanding and abilities to lead and train others.
	The focus is on strategic action.
\end{itemize}

\subsubsection{Target Audience:}
Target audience describes the audience, which is targeted by the training. This can be condensed to the following.
\begin{itemize}
	\item \textbf{Student/Trainee:} Student and trainees have little to none practical knowledge. Training can be used for students and trainees, to enhance their knowledge and to practice theoretical courses, see ~\cite{10.1145/3197091.3197123,10.1145/3159450.3159591,7943509,lehto}.
	\item \textbf{IT User:} IT users use the IT but have little to none knowledge about IT security. Users can get trained to understand principles of IT security and to grow awareness.
	\item \textbf{IT Professional:} Professionals have little to medium knowledge about IT security. Their professional focus is in specific sectors, therefore, they receive IT security knowledge for their sectors. 
	\item \textbf{IT Specialist:} Specialists already have a comprehensive knowledge in IT security. Therefore, the training is focussed on specific aspects.
	\item \textbf{Management:} Management has little knowledge about IT security, but a broad overview. By the training, management can understand changed settings better.
\end{itemize}

\subsection{Training Environment}

The training environment details the environment around the training, consisting of training type and scenario. Both characteristics are described in the following.

\subsubsection{Training Type:}
\input{textblocks/training_type}

\subsubsection{Scenario:}

The scenario is a main component of cybersecurity training. Scenarios are needed to reach the goal of the training and are described by the following characteristics.

\begin{itemize}
	\item \textbf{Supervision:} Either the training is supervised or unsupervised. Cyber range trainings are typically supervised, while jeopardy CTFs are unsupervised.
	\item \textbf{Style:} The style describes how the different challenges within the training are setup. Free-/Multi Choice can be the case with CTFs. Other directions are problem-driven and storyline-driven, if the challenges are arranged around a problem or a central storyline.
	\item \textbf{Challenges:} The challenges are the content of the training. These are defined by target and type. The target of the training can be a network, host, application, protocol, data, person, or physical. For solving the challenge types, foot-printing, scanning, enumeration, exploitation, pivoting, privilege escalation, covering tracks, and maintaining access may be needed. Souissi~\cite{7293480} and Lehto~\cite{lehto} use similar characterisations for attacks respectively competence areas.
\end{itemize}

\subsection{Training Setup}
The training setup further describes the training itself with the scoring, roles, the training mode as well as the customization level.

\subsubsection{Scoring:}
\input{textblocks/scoring}

\subsubsection{Roles:}
%The exercise planning process determines the participants, exercise scenario, injects and the execution order for the course of the exercise.
\input{textblocks/roles}

\subsubsection{Training Mode:}
\input{textblocks/training_mode}

\subsubsection{Customization Level:}
\input{textblocks/customization}
\section{\uppercase{Examples and Case Studies of Cyber Training and Education Systems}}\label{sec:examples}
\input{textblocks/case_study}
\input{textblocks/conclusion}


\bibliographystyle{splncs04}
\bibliography{CyberTaxi}
\end{document}

%% file: textblocks/introduction.tex
\section{\uppercase{Introduction}}\label{sec:introduction}% done Peter
%Motivation/Bedrohung/Ausgangslage
With increasing digitalization and the integration of computers into the daily life, the number of threats is also rising.
The amount of sophisticated attacks is increasing every year and poses a challenge in terms of efficient detection and countermeasures.
At the same time, the required technical knowledge of an intruder decreases with the development of more automated tools~\cite{Lipson2002,CERTDivision2003}.
The defence against these attacks is based on the continuous use and monitoring of security tools.
Well-trained personnel is required for this. 
However, current statistics show a global shortage of approximately four million information security professionals~\cite{ISC2}.
In recent years, many different training systems have been developed with focus on cybersecurity.
Companies﻿ have ﻿emerged ﻿offering ﻿new ﻿certifications ﻿and ﻿universities ﻿are ﻿developing ﻿cybersecurity ﻿degree﻿ programs.
Interactive training systems are further approaches that follow our motivation as cybersecurity trainers: Theoretical knowledge is good, practical proficiency is better.
These systems utilize gamification and playful scenarios to train participants in specific topics~\cite{Amorim2013}.

Interactive training systems offer the opportunity of real-time feedback and specific education.
The focus is on instructing the participants with an efficient method and the desirable knowledge. 
%In addition, these systems are ideal for recruiting and credit. 
These systems create awareness for security threats and the resulting impact in real life. 
In order to provide an overview of a large range of practical possibilities, we present a taxonomy to structure interactive cyber training and education.
This overview allows universities, companies, and other institutions to 
\begin{itemize}
\item identify gaps in their training system.
\item improve existing training systems and intensify the training. 
\item provide a guideline for establishing new training possibilities.
\end{itemize}
% Beside this, we provide an approach for performance-based learning and assessment.

% Our purpose is to train people's skill to solve problems in changing situations by applying acquired knowledge, ability, techniques, and rules.
The taxonomy includes surrounding aspects, e.g., training of teamwork, communication capabilities as well as reporting processes.
As far as we know, there is no holistic taxonomy for practical and interactive cyber training and education.
Such a taxonomy is mandatory to support education, structure the entire area, and accelerate further research.
It enables a comprehensive evaluation of interactive training systems and the visualization of the requirements gap.

%Aufbau des Papers \\
The paper is structured as follows:
Section \ref{sec:requirements} establishes the requirements for a taxonomy.
In the following section, we briefly discuss related approaches with their advantages and disadvantages.
In Section \ref{sec:taxonomy}, we describe our taxonomy in detail.
Afterwards, we classify other examples with our taxonomy and show the practical benefit.
Beside this, we discuss the presented aspects on our case study.
Finally, we conclude the paper and give future directions.
%Cyber Defence Training Needs Analysis

%% file: textblocks/requirements.tex
\section{\uppercase{Definition of Terms and Requirements}}\label{sec:requirements}% done Peter
%Anforderungen definieren\\
%Was soll in der Umgebung alles ausführbar sein\\
%Definition ICTE
We define \textit{Interactive Cyber Training and Education} (ICTE) as follows: 
%\frqq{}
%\glqq

\vspace{5pt}
\noindent\textbf{Definition:} 
ICTE is a comprehensive set of hands-on approaches in a secure and observable environment that enables participants to become engaged in learning and practice their cyber skills and to acquire new skills.

%\grqq
%\flqq
\vspace{5pt}
% Was ist eine Taxonomy
For a better understanding of the following taxonomy, a common basis of definition is mandatory.
In general, a taxonomy structures a knowledge field to provide an overview about a specific area and its possibilities.
It divides the topic hierarchical into main groups and subcategories.
A taxonomy should focus on the following, ideal properties~\cite{Landwehr1994,Lindqvist1997}:
\begin{itemize}
\item The categories have to be mutually exclusive, i.e., no overlapping between the categories.
\item Clear and unambiguous classification criteria. %(a repeated classification must produce the same results)
\item Comprehensible and useful as well as comply with established terminology.\\
\end{itemize}

% Anforderungen an die Taxonomy und die Ausbildungsumgebung
Beside the requirements for the taxonomy of ICTE, there are the following demands on systems:
\begin{itemize}
\item What are the optimal approaches and motivation for developing new cyber skills? %motivation
\item What skills and competencies in security are required to move to a more proactive position?
\item Which kind of training system requires which functionalities and possibilities?
\item What changes in terms of process, technology, and staff are required in the operational environment to support new abilities?
%\item What are the main cyber security concerns?
\item What are the business objectives and strategic goals of an organization from a security point of view?\\
%\item What should be executable in the playground environment? % Website, Jeopardy, Attack-Defence
%\item Which kind of training system requires which specific secured environments? % protected and secure environment
%\item What connections and functionalities should be available?\\ % Verfügbare Funktionalitäten in der Umgebung - Internet, Intranet
\end{itemize}

%Comparable criteria: Beside the requirements of a taxonomy
Furthermore, there are the following assessment criteria to further evaluate a classified training system.
These criteria follow the National Institute of Standards and Technology (NIST)~\cite{AcademicExcellence2017}, which developed the cybersecurity workforce framework for the National Initiative for Cybersecurity Education (NICE).
The requirements are tantamount to the ISO/IEC 25010~\cite{ISO/IEC2011} and ISO 9126~\cite{ISO9126} standards, which focus more on system and software quality properties.
These are also valid to our scenario.
\begin{itemize}
	\item \textbf{Functional Completeness:} The degree of realization to cover all the specified tasks and user objectives.
	\item \textbf{Functional Correctness:} The degree to which the cyber training system provides the correct and reproducible results with the needed degree of precision.
	\item \textbf{Learnability:} Efficiency to achieve the specified goal of learning individual and as cooperative team.%(Effectivness, Efficiency)
	\item \textbf{Operability:} The difficulty to run the training.% (Operate and Control)
	\item \textbf{Accessibility:} Amount of expertise to be successful in a scenario.
	\item \textbf{Adaptability:} The degree to customize the scenarios to the expertise of the participants.
	\item \textbf{Portability:} The degree to transfer scenarios from an education system to another.
	\item \textbf{Maintainability:} Difficulty to operate the system and to fulfill a training session.
	\item \textbf{Modularity:} Possibility to adapt and extend the system to current needs.\\
\end{itemize}

%% file: textblocks/sota.tex
\section{\uppercase{Related Work}}\label{sec:relatedwork}

Several different taxonomies were developed, e.g., taxonomies for \textit{Computer System Attack Classification}~\cite{Amoroso1994,Paulauskas2006,Neumann,6113347,10.5555/240069,brno,hansman}. 
These list a comprehensive set of attacks and focus on a structural overview of attacks.  
Jouini et al.~\cite{JOUINI2014489} classify security threats in information systems by threat source, agent, motivation, intention, and impact. 
Easttom and Butler~\cite{8666559} describe a taxonomy of cyber attacks based on a modified McCumber cubes. 
Amongst others, they classify the categories transmission, storage, technology, policy and practices, education, training, and awareness. 
Simmons et al.~\cite{avoidit} propose a taxonomy of cyber attacks called AVOIDIT, classifying attack vector, operational impact, defence, informational impact, and target. 
All these taxonomies can be used to develop security challenges, such that a wide spectrum of knowledge is necessary to solve them.

Different approaches focus on \textit{Cyber Range Training}. 
ECSO explains cyber ranges in their WG5 Paper~\cite{ecso}, but neither a definition nor a taxonomy is given. 
Priyadarshini~\cite{phdthesis} analyses and classifies existing cyber ranges based on infrastructure association, cloud usage, teams, and deployment. 
Other aspects are left out. 
Yamin et al.~\cite{YAMIN2020101636} build a taxonomy based on literature review. 
The taxonomy includes scenarios, monitoring, learning, management, teaming, and environment. 
Based on the taxonomy the authors describe several tools used within cyber ranges. 
The taxonomy is specific for cyber ranges and does not take target audience, proficiency level, and scoring, amongst others, into account.

Several papers describe \textit{Cybersecurity Exercises}. 
INCIBE~\cite{Diez2015} analyses cyber exercises and builds a short taxonomy based on the factors focus, model, vertical sector, scope for participation, and dissemination of results. 
It has a rough structure and a high level focus. 
The perspective is only for coverage of subject areas and educational view. 
Beyer and Brummel~\cite{beyer2015implementing} describe different factors for effective cybersecurity training. 
Kick~\cite{Kick2014} depicts playbooks in detail. 
Others analyze different environments for cyber training~\cite{8166396,205231,kypo,beuran,183455}. 
Different papers relate to aspects of gamification, serious games, and education~\cite{10.1145/3197091.3197123,10.1145/3159450.3159591,7943509,lehto}. 
As a result, a holistic taxonomy for the management and organization of ICTE is missing.

%% file: textblocks/technical_setup.tex
\subsection{Technical Setup}
The technical setup consists of environment structure, deployment, and orchestration.
These are described in detail in the following.

% done Peter
\subsubsection{Environment Structure:} 
The environment structure refers to the basic characteristic of the event.
This characteristic is composed of the following sub-characteristics:
\begin{itemize}
\item \textbf{Tabletop Style:} A session that involves the movement of counters or other objects round a board or on a flat surface.%)board games, card games, dice games, miniature wargames or tile-based games)
\item \textbf{Online Platform:} The digital service describes a wide range of interactive possibilities available on the internet including marketplaces, search engines, social media, creative content outlets, app stores, communications services, payment systems, services comprising the collaborative economy.
\begin{itemize}
\item \textbf{Collaboration Platform:} The environment allows organizations to incorporate real-time communication capabilities and providing remote access to other systems. This includes the exchange of files and messages in text, audio, and video formats between different computers or users.
% While products can be goods or services, platforms are access gateways
\item \textbf{E-Learning Platform:} A software application for the administration, documentation, tracking, reporting, and delivery of educational courses, training programs, or learning and development programs.
\end{itemize}
\item \textbf{Hosting:} A cyber training based on single hosts uses primarily a personal computer to providing tasks and challenges for a user. It allows a direct interaction with the systems. 
\item \textbf{Network Infrastructure:} Dependent of the realization type - simulated, emulated, or real - a network-based environment consists of servers and clients, which are connected to each other in a local area network (LAN) or wide area network (WAN).
%\todo{ggf. weglassen bei Platzmangel}
\begin{itemize}
	\item \textbf{Real:} Physical components are used to connect the systems and to setup a scenario.
	\item \textbf{Simulated:} A simulation copies the network components from the real world into a virtual environment. It provides an idea about how something works. It simulates the basic behavior but does not necessarily abide to all the rules of the real systems.
	\item \textbf{Emulated:} An emulator duplicates things exactly as they exist in real life. The emulation is effectively a complete imitation of the real thing. It operates in a virtual environment instead of the real world.
\end{itemize}
\end{itemize}

%\subsubsection{Infrastructure}
%Monitoring\\
%Flexibel Erweiterbar und Anpassbar\\
%Automatisierung bei der Erstellung?!\\

\subsubsection{Deployment:}
The environment of cyber training can either be deployed on premise or on cloud infrastructures, as shown in the following.

\begin{itemize}
\item \textbf{On Premise:} The environment for the training can either run on physical or virtual machines. Either way, the data is stored locally and not on cloud; nor is a third party involved. The benefit of virtual machines is the maximum of configurability. The advantages of on premise solutions are the physical accessibility, which makes it possible to use the complete range of cyber challenges.
\item \textbf{Cloud:} A training setup deployed in the cloud has on-demand availability of computer system resources, especially data storage and computing power, without direct active management by the user. In contrast to on premise setups, cloud solutions are rapid elastic on request. So the training can be adapted flexible on a large amount of users and is easily usable world wide. %\cite{NIST CLOUD DEFINITION FRAMEWORK}
%The environment for the training can be deployed in the cloud. While the data lies somewhere else, machines can be added dynamically.
\end{itemize}

\subsubsection{Orchestration:} % done Peter
We understand orchestration as the composition of parts and components of a pool of tasks. The goal is to setup a holistic scenario and integrate cyber training session.
Furthermore, it includes a declarative description of the overall process in the form of a composite and harmonic collaboration.
A system typically exists of functions, processes, and data. It provides a common service embedded in an environment for the specified purpose.
Beside this, orchestration has also a strong relation to the deployment strategy and the customization possibilities.
Well known approaches are tools like Chef \cite{chef}, Puppet \cite{puppet}, Ansible \cite{ansible}, and SaltStack \cite{saltstack}.

A flexible concept for orchestration and maintainability is important for the administration of the cyber training system.
Especially, the possibility of a fast troubleshooting in case of live events is mandatory.
Nevertheless, it also has an impact on the participants in relation to user experience and quality of service in providing a training with adequate atmosphere.

The criterion orchestration is further divided in the degree of process automation, portability, maintainability, and compatibility.
\begin{itemize}
	\item \textbf{Automation:} It specifies the automation of processes and the amount of human interaction with the system to maintain and administrate, especially for repetitive exercise.
	Subclasses for automation are non-automation, partially-automation, and full-automation.
	\item \textbf{Portability:} The possibility to exchange data, challenges, or entire scenarios to other environments or locations.
	The portability can be separated in miscellaneous approaches and the usage of common data format for exchange like YAML, Extensible Markup Language (XML), and JavaScript Object Notation (JSON) \cite{DBLP:conf/im/SteinbergerSGB15,Howard1998}. %Koch2013
	%SNMP CIDF IDMEF IODE FFINE IDXP
	The objective of future research direction is an exchange format for entire cyber scenarios to flexible deploy these in different environments and locations.
	\item \textbf{Maintainability:} Maintainability represents effectiveness and efficiency with which a session can be modified or adapted to changes.
	A modular concept has advantages in reusability and combinability.
	\item \textbf{Compatibility:} The Compatibility deals with the technical interaction possibilities via interfaces to other applications, data, and protocols.
\end{itemize}

%% file: textblocks/training_type.tex
Education in cybersecurity follows different approaches \cite{yurcik2001different}.
The level of interaction and hands-on experience distinguishes different types of training. 
For interactive cyber training, the following training types exist.

\begin{itemize}
\item \textbf{Table Top:} 
This type is a lightweight, but intellectually intense exercise.
In this setting, the involved teams or participants focus on opposing missions.
On a theoretical basis, the teams develop different strategies and countermeasures to explore the offensive cyber effects on operations. 
Table top trainings could be based on speech-only, text-only, or multimedia
\cite{DoDCTT,Kick2014}.
%\item Tutorial Approach: 
%This training type consists mainly of self-learning. 
%It utilizes self-study texts and learning platforms on various topics inn cyber security. \cite{yurcik2001different}
% \todo[inline]{Kann rausgelassen werden, da nicht interaktiv}
\item \textbf{Project Approach:} 
In this type of training, hands-on projects are to be completed during the training.
Thereby, the participants learn and understand the basic concepts of security. 
During the projects, the teachers can intervene and control the learning process \cite{yurcik2001different}.
\item \textbf{Capture the Flag:}
Capture the Flag (CTF) is a well-known cybersecurity contest in which participants compete in real-time.
Several distinct kinds of CTF have evolved in the recent years, including quiz, jeopardy, attack-only, defence-only, and attack-defence~\cite{davis2014fun}.
\item \textbf{Cyber Training Range:}
A cyber range provides an environment to practice network operation skills.
It should represent real-world scenarios and offer isolation from other networks to contain malicious activity.
In this training type, complex attacks take place in a simulated environment.
The participants perform divers educational hands-on activities according to their role. 
Possible trainings are classroom practice, single team, and multiple team trainings.
In these trainings the roles that are not covered by participants are simulated or covered by the instructors \cite{davis2013survey,kypo}.

\end{itemize}

%% file: textblocks/scoring.tex
Scoring is an important component of a cyber training.
Depending on the purpose of the training, the scoring provides means to motivate the participants and a way to give feedback.
It is also used to track the progress during a training.
For competition-oriented trainings, like CTFs, a scoring is necessary.
The scoring can be based, but is not limited to monitoring systems, defined objectives, or over-the-shoulder evaluation mechanisms.

\begin{itemize}
\item \textbf{Awarding:} In this variant of scoring, participants get awards for predefined actions or achievements.
These awards can be granted both manually and automatically.
Furthermore, a mixed approach is possible, e.g., by automatically giving awards for general objectives and manually giving awards for outstanding achievements.
In general, awarding has a lower granularity than the detailed assessment and requires less administrative effort, but gives reasonable feedback and motivation for the participants.
\item \textbf{Assessment:} This scoring variant is more complex than awarding and allows to assess participants and compare them to each other.
The assessment scores can be assigned in different ways. 
One type is the static setting of different scores for tasks and objectives. 
In order to distinguish it from awarding, the degree of difficulty can be included here. 
Furthermore, the scores for different tasks can be set dynamically using mathematical functions. 
But also other dynamic methods, such as the Elo Rating System \cite{pelanek2016applications}, are covered by this variant.

\item \textbf{No Scoring:} Depending on the training, a scoring is not necessarily needed.
\end{itemize}

%% file: textblocks/roles.tex
\label{subsubsec_roles}
Participants in a training are split in different teams, according to their skills, role, and tasks during a training. 
For the identification, each team has a color assigned based on its role.
The following teams are commonly used in cyber trainings and exercises \cite{Kick2014,vykopal2017lessons}.

\begin{itemize}
\item \textbf{Green Team:}
The operators that are responsible for the exercise infrastructure build this team. 
Before a training, this team sets up and configures the environment and takes it down afterwards.
During a training, it also monitors the environments health and handles problems that may arise. 
\item \textbf{White Team:} 
This team consists of instructors, referees, organizers, and training managers. 
They design the training scenario including objectives, rules, background story, and tasks. 
During the training, this team controls the progress and assigns tasks to the teams. 
These so-called \textit{injects} also include simulated media, operation coordination, or law enforcement agencies. 
Giving hints for the training teams could also be part of this team.
\item \textbf{Red Team:}
This team consists of people authorized and organized to model security adversaries.
They are responsible to identify and exploit potential vulnerabilities present in the training environment.
Depending on the training environment, the tasks can follow a predefined attack path.
\item \textbf{Blue Team:} 
The group of individuals that is responsible for defending the training environment.
They deal with the red team's attacks and secure the compromised networks.
Guidelines for that team are the training rules and local cyber law.
\end{itemize}

Additionally, there are further roles involved in training, which are summarized in the following teams \cite{Wright_blackhat_17,YAMIN2020101636}.

\begin{itemize}
\item \textbf{Transparent Team:} Members of this team observe the training.
Usually, these people have a defined purpose, but have no influence on the training itself.
Possible purposes are learning about the training topic and roles, studying strategies of participants, or supervising employees.
\item \textbf{Yellow Team:} Members of this team perform not only tasks like generating legitimate network traffic and user behavior but also perform erroneous actions that lead to vulnerabilities and attacks.
This team can also include the regular system builders, like programmers, developers, and software engineers and architects.  
\item \textbf{Purple Team:} In a training, this team is a bridge between red and blue teams that helps to improve the performance of both.
Through joint red-blue activities it improves the scope of the training participants.
Goals are to maximize the Blue Teams capability and the effectiveness of Red Teams activities.  
\item \textbf{Gray Team:} Bystanders of a training form this team.
They do not necessarily have a specific intention or purpose, but an interest in the training event itself.
It is also possible that this team interacts with participants and thereby unintentionally influences the training.
\item \textbf{No specific role:} Individuals who do not fit into the defined teams can be assigned to this role. % Observer?
\end{itemize}

According to \cite{Wright_blackhat_17}, the orange team completes the so called Color Wheel.
This team is of special importance in a holistic system development process.

%%%
%%%vykopal2017lessons
%%%Wright_blackhat_17
%%%YAMIN2020101636
%%%
%%%\todo[inline]{Cyber Exercise Playbook: Jason Kick - https://www.mitre.org/publications/technical-papers/cyber-exercise-playbook}
%%%Organizers: Steps for a successful\\
%%%\todo[inline]{https://www.enisa.europa.eu/events/cyber-exercise-conference/presentations/10. Conf Paris -June 2012- - P. TRIMINTZIOS-ENISA.pdf}
%%%
%%%\todo[inline]{https://www.nist.gov/sites/default/files/documents/2017/05/23/cyber\_ranges\_2017.pdf}

%% file: textblocks/training_mode.tex
Training mode defines the mode in which the training is accomplished. 
The training mode has three different alignments.
\begin{itemize}
\item \textbf{Single:} A single player plays against others. Others can be real persons, but also scripted opponents.
\item \textbf{Team:} A team plays against others. In this alignments, each player can bring its expertise into the training, focussing on different aspects. Examples are Blue and Red Teams.
\item \textbf{Group:} A group plays against others. In this setting, the group members might not know each other. Example are CTF competitions and training for the entire organization in a breach scenario.
\end{itemize}

%% file: textblocks/customization.tex
Depending on the goal of the training, the training setup can be customized.
A distinction is made here between three variants.
\begin{itemize}
\item \textbf{General:} A general purpose training setup is not, or only little customized.
This variant is suited for an entry level training or to learn about general processes without regard to the underlying setup. 
\item \textbf{Specific:} The training setup can be customized for a specific training goal or target audience.
Examples for this variant are specific trainings within the High School education \cite{hembroff2015development} or for the health sector \cite{rajamaki2018cybersecurity}.
\item \textbf{Individual:}
The most tailored variant is an individual customization.
Hereby, the training setup corresponds to a real environment in the best possible way.
Exemplary uses of this variant are the training of teams in their environment or the training of new expert-level employees.
\end{itemize}

%% file: textblocks/case_study.tex
%Beschreibung unserer CTF Erfahrung\\
%\todo[inline]{Class Capture-the-Flag-Exercises: Jelena Mirkovic, Peter Peterson} 

A tailored modification of a CTF for educational purposes is described by ~\cite{183455}, named Class CTF (CCTF). 
%They propose a class CTF for educational purposes. 
The idea behind this approach is to maximize the learning outcome and minimize the time spent for a CTF event. 
They observed a higher motivation of their students helping each other solving the challenges. 
Also much more interest in learning and practicing skills during hands-on exercises was noticed.
This seems to be a usual behavior if changing the learning setup to more hands-on exercises. 
Nevertheless, this approach can be categorized in our taxonomy as follows: The training setup is based on Red and Blue team roles. 
To give all participants the same chances, the presented scoring is based on fixed solutions for every challenge and, therefore, static. 
CCTFs follow a team based approach and are designed for students in universities to improve their skills and foster communication and collaboration abilities. 
The proficiency level starts from beginner level and can evolve during a series of CCTFs. 

The presented taxonomy progressed during our own Capture the Flag experiences. 
Whereas this taxonomy was very basic during the first phase of our initial CTF in 2015, it evolved further during the next events. 
Table \ref{tab:tableevents} summarizes our past events with some details.
In the next section, our CTFs are categorized based on the taxonomy. 
This is followed by our experiences during the organization of various CTF events.
%\todo[inline]{table mit CTFs? inkl. Statistiken}
\begin{figure*}[b] 
	\centering
	\noindent
\small{	
\begin{tabular}{c|p{24mm}cccp{26mm}p{22mm}}
	Year & Title & Duration & Teams & Attendees & Challenges & Tracks \\ 	
	\hline 
	2015 & The Beginning & 9 h & 11 & 31 & 34 + 1 Easteregg & Beginner,\newline Advanced \\ 
	2016 & A New Hope & 8 h & 14 & 49 & 18 + 2 Eastereggs & Beginner,\newline Advanced,\newline Professional  \\ 
	2017 & 24:\newline The Revolution & 24 h  & 18 & 69 & 48 + 2 Eastereggs & Beginner,\newline Advanced \\ 
	2018 & Dark Fiber & 18 h & 24 & 92 & 39 + 3 Eastereggs,\newline 3 Scenarios/Maps & Jeopardy,\newline Attack-Defence \\ 
	2019 & The 5th Element & 18 h & 29 & 129 & 50 + 2 Eastereggs & Jeopardy \\ 
\end{tabular}}
	\caption{Overview of accomplished CTF events. Each session consisted of an additional Online-Qualifying over one week.}
	\label{tab:tableevents}
	
\end{figure*} 
%\todo[inline]{Tabelle referenzieren}
% Jahr - Name - Zeitansatz - Teilnehmer - Challenges - Level - Modi

\subsection{Application of the Taxonomy}

 Since our goal was to organize an event for the students of our university, we followed an approach to support their skills, their team spirit, and nevertheless fun during the event. That said we classify our CTFs based on our \textit{taxonomy} as follows. 

\begin{itemize}
\item \textbf{Audience}\
\begin{itemize}
\item \textbf{Target Audience:} Students and IT professionals.
\item \textbf{Sector:} Public (NGOs, government agencies), and academic.
\item \textbf{Proficiency Level:} Beginner up to expert level.
\item \textbf{Purpose:} Improve skills and collaboration abilities within the teams.
\end{itemize} 
\item \textbf{Training Environment} \
\begin{itemize}
\item \textbf{Training Type:} Capture the flag including quiz and jeopardy. In 2019, we also set up an attack-defence track. 
\item \textbf{Scenario:} Storyline-driven including some free-/multiple choice questions..
\end{itemize}
\item \textbf{Training Setup} \
\begin{itemize}
\item \textbf{Scoring:} Assessment, as it uses dynamic scoring.
\item \textbf{Roles:} No specific roles.
\item \textbf{Training Mode:} Team.
\item \textbf{Customization Level:} Individual, because the incentive is to provide an environment near to real world scenarios.
\end{itemize}
\item \textbf{Technical Setup} \
\begin{itemize}
\item \textbf{Environment Structure:} Online platform in sense of E-Learning platform, and hosting.
\item \textbf{Deployment:} On-premise.
\item \textbf{Orchestration:} Partial degree of automation and modular approach for designing the services. 
\end{itemize}  
\end{itemize}

\subsection{Phase 1: Organization and Development}\label{sec:casestudy}
%Wie verläuft das Management im Vornhinein, wie läuft das Management während der Veranstaltung...regelmäßige meetings, ticketsystem, versionsverwaltung, klare verantwortlichkeiten\\
% https://blog.legitbs.net/2017/12/finals-building-def-con-ctf.html
% https://blog.legitbs.net/2017/10/hype-meetings-and-workflows-building.html
%Lessons Learned\\
%Aus Zusammenfassungen der CTFs/Statistiken rausziehen, was für ein Event, Teilnehmer, Dauer etc. 
% ctf poster, kategorie, auslastung cluster, ressourcen vgl. attack defense?

In the first step, it starts with an idea to organize a CTF event. 
Meetings are scheduled and all participants are on a hype that the event will be a great success. 
But latest during the first crunch-time, the hype ends and all participants realize the hard work. 
Therefore, it is necessary to structure and organize the phases. 
Choose a project management that fits for the needs. 
It is important to have regular meetings, but also a ticketing system and version control for the challenges and underlying infrastructure.

During the first meetings, the audience for the event and the requirements need to be defined. 
We used our \textit{taxonomy}, as shown in the previous section, to help ourselves. 
The classification of a planned event allows organizers to derive additional constraints, e.g., to support multiple proficiency levels based on categories or changing the level of difficulty based on successful solves of challenges.
We followed an approach to support their skills, their team spirit, and nevertheless fun during the event. 
This includes a story-driven scenario including some free-/multiple choice questions, e.g., eastereggs. 
Thereby, we develop several challenges for beginner as well as experts. 
In order to provide equality and increase the motivation, we provide different tracks with separate scoring based on the proficiency level. 
We have learned from experience that dynamic scoring makes it easier to determine the individual difficulty of the various challenges. 
As a result, the training setup includes our own developed scoring system. 
We promote no specific roles in our setup, but in the last years some teams joined the event with observer participants. 

As an advice from our own experiences during the development of an event, we recommend planning sufficient buffers for unexpected issues regarding challenges and infrastructure. 
This saves the organizers from excessive crunch-time.  
Our technical setup was complex so far. 
It includes real network infrastructure and on-premise hardware for hosting the environment. 
On top of these hardware machines, we setup a virtual environment consisting of virtual machines and virtual networking infrastructure comparable to a real data centre. 
To minimize the workload during the event, we tried to orchestrate availability checks and restarts of vital services. 

% \texttt{Audience: -> Target Audience: Student/Trainee; -> Sector -> Academic: School; -> Propose: Skills,Collaboration; Proficiency Level: Beginner, Professional; Training Environment: -> Training Type: -> Capture the Flag: Quiz, Jeopardy, Attack-Defence(first in 2018); Training Setup: -> Scoring -> Assessment: Dynamic; -> Training Mode: Team; Technical Setup: -> Environment Structure: -> Online Platform: E-Learning; -> Hosting; -> Deployment: -> On-premise: Virtual} \\
%\begin{itemize}
	%\item Planung CTF anhang der Taxonomy (Audience - Target Student; Sector, Academic - University; Type CTF, Jeopardy; Proficiency Beginner, Professional, Auswahl Environment Hosting, Scoring?) % Schwierigkeitsgrad durch den Nutzer bestimmt, anhand der solves
%	\item Zeitrahmen -> Anzahl Challenges, Training Mode -> Team
%	\item Vorbereitung / Entwicklung der Challenges ist ausschlaggebend für Erfolg, hoher Zeitinvest, der je nach Audience eine Wiederverwendung ermöglicht
%\end{itemize}
% crunchtime / buffer zur vermeidung von crunchtime
\subsection{Phase 2: Testing and Dry Run}
%The dry run is a complete test of the proposed cyber exercise to get diverse feedback on it.
The main part of the second phase is testing. 
During this phase, all challenges are checked by an individual or small team independent from the developers of a challenge. 
If the event is based on a story line, all transitions to following challenges and activations of challenges after a successful solve need to be checked.
 Furthermore, it is necessary to keep in mind that nothing is more frustrating for participants than investing a lot of time in solving a challenge that is buggy or not solvable. 
 Therefore, testing is more than essential for a successful event. 
%During our experience in developing challenges we observed various times that a developers of challenges rate their challenges as easy to solve whereas tester rate the same challenge as complex. 
The testing phase is also a good point in time to check the overall progress in developing challenges. 
To support the participants during the event, walk-throughs should be developed and tested in parallel. 
If a hint system is planned for the main event, it should also be checked during this phase.
Nevertheless, every hint system has its own pros and cons. 
We did not find a satisfactory hint system for all participants so far; except for providing no hints for a single team. 
Finally in this phase, check the infrastructure readiness for the number of planned participants.
%\begin{itemize}
%	\item Test der einzelnen Challenges
%	\item bei Storyline Test der Challenges und Überleitungen/Freischaltung der Challenges basierend auf Lösung/Zeit
%	\item Lösbarkeit der Challenges prüfen, nichts ist frustriender als wenn eine Chall nicht funzt und die Teiln. laaange dran sitzen
%	\item bei Hint System prüfen, ob/wie die Hints helfen 
%	\item Walktrough ggf. für Unterstützung während der Veranstaltung wichtig
%\end{itemize}
\subsection{Phase 3: Accomplishment}
The accomplishment phase usually takes place at two points in time: (1) qualifying and (2) main event. 
During the qualifying event, it is possible to gain experiences for the main event. 
Are the challenges too easy to solve? 
Is it necessary to split the participants into different tracks based on their skills? 
Are infrastructure resources planned adequately? 
Did the challenges need a supervision or are they robust enough? 
Taking these possible questions into account, the setup can be adjusted where appropriate. 
In case of different categories based on proficiency level, the participants should decide which track to choose. 
Additionally, it is important to plan enough resources in infrastructure and staff for the main event. 
It is usual that not all challenges and storylines run as expected. 
Preparation should also include unforeseen events such as network and power failures, unavailable challenges or challenges that cannot be solved even if tested in advance.

The event should start with a short introduction about the "Do's and Don'ts" during the event. 
All participant need to understand that any attempt to break the infrastructure or manipulation of the scoring system leads to disqualification. 
Have a system in place to monitor such attempts. 
If photos are taken during the event or names of persons or teams are published afterwards, the participants have to be asked for consent. 
Consideration should be given in advance to where this information will be placed afterwards.
This is necessary for the terms under which it is allowed to do so, keeping in mind regulatory requirements.  
Furthermore, a monitoring system should be in place to track the availability of all challenges and scoring systems. 
Necessary services should restart automatically in case of a failure as there is no time for manual troubleshooting during the event.

Regarding the IT security of the participants and infrastructure, the "Do's and Don'ts" are presented in the first step. 
Yet, various technical measures are feasible to prevent, e.g., manipulation of virtual machines, containers or network infrastructure. 
Virtual Machines (VMs) virtualize an underlying computer, whereas with containers the operating system is virtualized. 
Each way has consequences for the security, resources, reproducibility of exploits, and permissions. 
If a participant is able to break out of a docker container, other challenges could be manipulated in other containers or directly on the VM. 
Furthermore, if a participant is able to break out the virtual machine, also other VMs on the host or even the hypervisor can be manipulated. 
Therefore, it is vital to make even a short risk assessment of your infrastructure hosting the challenges. 
If a team is manipulating their own challenges and infrastructure, it might be acceptable to disqualify the whole team. 
However, if it cannot be determined who has manipulated a challenge or infrastructure, it may be necessary to cancel the entire event.
That would be very frustrating for the other teams playing a fair game.

\subsection{Phase 4: Cleanup and Maintenance}
The last phase starts with the collection of evaluation sheets handed out during the event. 
The feedback of the participants is a good measure to improve the quality of a CTF and at the same time input for further events. 
Dependent on the feedback form and participants, it includes feedback about the difficulty of the challenges, the setup, but also about the fringe. 
The traffic and log files generated during the event are a good starting point for further analytics. 
Outcomes could include information on team strategies, toolsets used, and additional solutions to your challenges
If desirable, initiate a call for write-ups and provide a platform for the teams to share their solutions. 
It can be inspiring to read write-up of different solutions. 
Furthermore, traffic collected during the event showing various kinds of attacks to provided services is very valuable for security research. 
To reuse the hosted challenges for further events, clean-up the provided machines afterwards. 
An easy solution is either using container or making a snapshot of all resources in advance. 
This enables the reset to this snapshot after the event. 
To keep the system save, shutdown all resources if they are not in use between events.

It must be ensured that all participants have given their consent for relevant information to be collected from evaluation sheets, log files, traffic, pictures, and team names. 
In case of collected data from a participant that did not gave the consent, these data need to be deleted immediately. 
Subsequently, the data should be prepared for publishing, depending on the needs. 
Additionally, sponsors definitely welcome a short report about the event. 
Further, if the event was a success, publishing the results gives the organization and development team a good standing in organizing cybersecurity events. 
%Auswertung der Evaluation und der Logs
%\begin{itemize}
%	\item Aufräumen der Systeme; Löschen / Bereinigen der Systeme die zur Verfügung gestellt wurden; ggf. für neues CTF "zurücksetzen" 
%	\item Auswertung Evaluation der Veranstaltung 
%	\item Auswetung Logfiles / Traffic kann helfen in der Forschung IT-Sec
%	\item Einwilligungserklärungen prüfen
%	\item Ergebnisse kommunizieren -> "Marketing" 
%	\item Report für Sponsoren
%	\item Zugänge deaktivieren/ Systeme bei Nichtgebrauch abschalten / kein Zugriff auf Systeme wenn keine VA läuft
%\end{itemize}

%% file: textblocks/conclusion.tex
\section{\uppercase{Conclusion and Future Work}}\label{sec:conclusion}
Cyber training and education systems are an important aspect in education of different persons in each sector. 
Interactive training can help to improve the security knowledge in a practical way. 
The development of such systems as well as the extension and improvement can be hampered because of a missing general classification. 
This is especially the case for specific systems with focus on an individual use case.

To overcome this shortcoming, we developed a flexible taxonomy for ICTE systems. 
The taxonomy provides a detailed description of all components with a focus on technical realization. 
All phases of the exercise life cycle are covered within the taxonomy to obtain a holistic approach. 
It supports the education and training of different roles under consideration of the current skill level.
This allows a targeted teaching in specific scenarios. 
In a next step, we showed examples of education and training systems, before we provided a case study based on conducted trainings. 
The trainings were categorized by our taxonomy. 
These examples further explain different training types.

In the future, we will conduct a survey about training systems, in order to apply the taxonomy to further systems.
For example, we will show the fully compatibility and extensibility with the NIST NICE Framework~\cite{Newhouse2017}.
This can help to enrich the taxonomy with more details.
Further non-technical possibilities are ICTE in context of assurances and certification levels. 
Beside this, the taxonomy motivates the discussion and highlights the advantages of such systems. 
In order to improve the trainings, we will compare different types of scoring methodologies and provide a better suited one. 
As different scenarios are developed, a universal scenario description format or language helps to compare and exchange scenarios. 
This will be designed in addition.

%% file: ms.bbl
\begin{thebibliography}{10}
\providecommand{\url}[1]{\texttt{#1}}
\providecommand{\urlprefix}{URL }
\providecommand{\doi}[1]{https://doi.org/#1}

\bibitem{Amorim2013}
Amorim, J.A., Hendrix, M., Andler, S.F., Gustavsson, P.M.: {Gamified Training
  for Cyber Defence: Methods and Automated Tools for Situation and Threat
  Assessment}. Nato Modelling \& Simulation Group (NMSG) Multi-Workshop,
  MSG-111  (2013)

\bibitem{Amoroso1994}
Amoroso, E.: {Fundamentals of Computer Security Technology}. Prentice-Hall
  (1994)

\bibitem{beuran}
Beuran, R., Pham, C., Tang, D., Chinen, K.i., Tan, Y., Shinoda, Y.:
  {Cybersecurity Education and Training Support System: CyRIS}. IEICE
  Transactions on Information and Systems  \textbf{E101.D},  740--749 (Mar
  2018)

\bibitem{beyer2015implementing}
Beyer, R.E., Brummel, B.: {Implementing Effective Cyber Security Training for
  End Users of Computer Networks}. SHRM-SIOP Science of HR Series: Promoting
  Evidence-Based HR  \textbf{3}(10), ~2018 (2015)

\bibitem{Bishop99whatdo}
Bishop, M.: What do we mean by ''computer security education''? In: 22nd
  National Information Systems Security Conference (1999)

\bibitem{CERTDivision2003}
{CERT Division}: {CERT Coordination Center - 2002 Annual Report}. Tech. rep.,
  Carnegie Mellon University, Software Engineering Institute (2003)

\bibitem{chef}
{Chef Software Inc.}: {Chef} (2020), \url{https://www.chef.io}, (Accessed on
  August 28, 2020)

\bibitem{concordia}
{CONCORDIA}: {Courses and Trainings for Professionals} (2020),
  \url{https://www.concordia-h2020.eu/map-courses-cyber-professionals/},
  (Accessed on August 28, 2020)

\bibitem{davis2014fun}
Davis, A., Leek, T., Zhivich, M., Gwinnup, K., Leonard, W.: {The Fun and Future
  of CTF}. In: 2014 {USENIX} Summit on Gaming, Games, and Gamification in
  Security Education (3GSE 14) (2014)

\bibitem{davis2013survey}
Davis, J., Magrath, S.: {A Survey of Cyber Ranges and Testbeds}. Tech. rep.,
  Defence Science and Technology Organisation Edinburgh (Australia) Cyber and
  Electronic Warfare Div (2013)

\bibitem{Diez2015}
D{\'i}ez, E.G., Pereira, D.F., Merino, M.A.L., Su{\'a}rez, H.R., Juan, D.B.:
  {Cyber exercises taxonomy}. {INCIBE}  (2015),
  \url{https://www.incibe.es/extfrontinteco/img/File/intecocert/EstudiosInformes/incibe_cyberexercises_taxonomy.pdf},
  (Accessed on August 28, 2020)

\bibitem{8666559}
{Easttom}, C., {Butler}, W.: {A Modified McCumber Cube as a Basis for a
  Taxonomy of Cyber Attacks}. In: {2019 IEEE 9th Annual Computing and
  Communication Workshop and Conference (CCWC)}. pp. 943--949 (2019)

\bibitem{ecso}
{European Cyber Security Organisation}: {WG5 Paper - Understanding Cyber
  Ranges: From Hype to Reality}. Tech. rep. (Mar 2020)

\bibitem{hansman}
Hansman, S., Hunt, R.: A taxonomy of network and computer attacks. Computers \&
  Security  \textbf{24},  31--43 (Feb 2005)

\bibitem{hembroff2015development}
Hembroff, G., Hanson, L., Vanwagner, T., Wambold, S., Wang, X.: {The
  Development of a Computer \& Network Security Education Interactive Gaming
  Architecture for High School Age Students}. {The USENIX Journal of Education
  in System Administration}  \textbf{25} (2015)

\bibitem{Howard1998}
Howard, J.D., Longstaff, T.A.: {A Common Language for ComputerSecurity
  Incidents}. Tech. rep., Sandia National Laboratories (1998)

\bibitem{ISC2}
{(ISC)2}: {Strategies for Building and Growing Strong Cybersecurity Teams}.
  Cybersecurity Workforce Study  (2019),
  \url{https://www.isc2.org/-/media/ISC2/Research/2019-Cybersecurity-Workforce-Study/ISC2-Cybersecurity-Workforce-Study-2019.ashx},
  (Accessed on August 28, 2020)

\bibitem{ISO9126}
{ISO/IEC}: {ISO/IEC 9126. Software engineering -- Product quality}. {ISO/IEC}
  (2001)

\bibitem{ISO/IEC2011}
{ISO/IEC 25010}: {ISO/IEC 25010:2011, Systems and software engineering —
  Systems and software Quality Requirements and Evaluation (SQuaRE) — System
  and software quality models}. {ISO/IEC} (2011)

\bibitem{10.1145/3159450.3159591}
Jin, G., Tu, M., Kim, T.H., Heffron, J., White, J.: {Game Based Cybersecurity
  Training for High School Students}. In: {Proceedings of the 49th ACM
  Technical Symposium on Computer Science Education}. pp. 68--73. SIGCSE ’18,
  Association for Computing Machinery, New York, NY, USA (2018)

\bibitem{JOUINI2014489}
Jouini, M., Rabai, L.B.A., Aissa, A.B.: {Classification of Security Threats in
  Information Systems}. The 5th International Conference on Ambient Systems,
  Networks and Technologies (ANT-2014), {Procedia Computer Science}
  \textbf{32},  489--496 (2014)

\bibitem{Kick2014}
Kick, J.: {Cyber Exercise Playbook}. {MITRE}  (2014),
  \url{https://www.mitre.org/sites/default/files/publications/pr_14-3929-cyber-exercise-playbook.pdf},
  (Accessed on August 28, 2020)

\bibitem{10.5555/240069}
Kumar, S.: {Classification and Detection of Computer Intrusions}. Ph.D. thesis,
  Purdue University, USA (1996)

\bibitem{Landwehr1994}
Landwehr, C.E., Bull, A.R., McDermott, J.P., Choi, W.S.: {A taxonomy of
  computer program security flaw}. ACM Computing Surveys  (1994)

\bibitem{lehto}
Lehto, M.: {Cyber Security Education and Research in the Finland's Universities
  and Universities of Applied Sciences}. {International Journal of Cyber
  Warfare and Terrorism}  \textbf{6},  15--31 (Apr 2016)

\bibitem{Lindqvist1997}
Lindqvist, U., Jonsson, E.: {How to Systematically Classify Computer Security
  Intrusions}. IEEE Symposium Security and Privacy pp. 154--163 (1997)

\bibitem{Lipson2002}
Lipson, H.F.: {Tracking and Tracing Cyber-Attacks: Technical Challenges and
  Global Policy Issues}. {Software Engineering Institute, CERT Coordination
  Center}  (2002)

\bibitem{183455}
Mirkovic, J., Peterson, P.A.H.: {Class Capture-the-Flag Exercises}. In: 2014
  {USENIX} Summit on Gaming, Games, and Gamification in Security Education
  (3GSE 14). {USENIX} Association, San Diego, CA (Aug 2014)

\bibitem{Neumann}
Neumann, P.G., Parker, D.B.: {A Summary of Computer Misuse Techniques}. In:
  12th National Computer Security Conference, Baltimore, MD. pp. 396--406 (Oct
  1989)

\bibitem{Newhouse2017}
Newhouse, W., Keith, S., Scribner, B., Witte, G.: National initiative for
  cybersecurity education (nice) cybersecurity workforce framework. NIST
  Special Publication 800-181  (2017)

\bibitem{AcademicExcellence2017}
Newhouse, W., Keith, S., Scribner, B., Witte, G.: {National Initiative for
  Cybersecurity Education (NICE), Cybersecurity Workforce Framework, NIST
  Special Publication 800 -181}. National Institute of Standards and
  Technology, US Department of Homeland Security, National Initiative for
  Cybersecurity Careers and Studies (NICCS)  (2017)

\bibitem{Paulauskas2006}
Paulauskas, N., Garsva, E.: {Computer System Attack Classification}. {IEEE
  Automation and Robotics}  (2006)

\bibitem{pelanek2016applications}
Pel{\'a}nek, R.: {Applications of the Elo rating system in adaptive educational
  systems}. Computers \& Education  \textbf{98},  169--179 (2016)

\bibitem{phdthesis}
Priyadarshini, I.: {Features and Architecture of The Modern Cyber Range: A
  Qualitative Analysis and Survey}. Ph.D. thesis, University of Delaware (2018)

\bibitem{puppet}
{Puppet, Inc.}: {puppet} (2020), \url{https://www.puppet.com}, (Accessed on
  August 28, 2020)

\bibitem{rajamaki2018cybersecurity}
Rajam{\"a}ki, J., Nevmerzhitskaya, J., Vir{\'a}g, C.: {Cybersecurity education
  and training in hospitals: Proactive resilience educational framework
  (Prosilience EF)}. In: 2018 IEEE Global Engineering Education Conference
  (EDUCON). pp. 2042--2046. IEEE (2018)

\bibitem{ansible}
{Red Hat, Inc.}: {Red Hat Ansible} (2020), \url{https://www.ansible.com},
  (Accessed on August 28, 2020)

\bibitem{saltstack}
{SaltStack, Inc.}: {Saltstack} (2020), \url{https://www.saltstack.com},
  (Accessed on August 28, 2020)

\bibitem{avoidit}
Simmons, C., Ellis, C., Shiva, S., Dasgupta, D., Wu, Q.: {AVOIDIT: A Cyber
  Attack Taxonomy}. In: 9th Annual Symposium on Information Assurance
  (ASIA’14). pp. 2--12 (2014)

\bibitem{7293480}
Souissi, S.: A novel response-oriented attack classification. In: 2015
  International Conference on Protocol Engineering (ICPE) and International
  Conference on New Technologies of Distributed Systems (NTDS). pp.~1--6 (2015)

\bibitem{DBLP:conf/im/SteinbergerSGB15}
Steinberger, J., Sperotto, A., Golling, M., Baier, H.: {How to exchange
  security events? Overview and evaluation of formats and protocols}. In:
  Badonnel, R., Xiao, J., Ata, S., Turck, F.D., Groza, V., dos Santos, C.R.P.
  (eds.) {IFIP/IEEE} International Symposium on Integrated Network Management,
  {IM} 2015. pp. 261--269. {IEEE} (2015)

\bibitem{8166396}
Suba{\c{s}}u, G., Ro{\c{s}}u, L., B{\u{a}}doi, I.: Modeling and simulation
  architecture for training in cyber defence education. In: 2017 9th
  International Conference on Electronics, Computers and Artificial
  Intelligence (ECAI). pp.~1--4 (2017)

\bibitem{10.1145/3197091.3197123}
{\v{S}}v{\'{a}}bensk{\'{y}}, V., Vykopal, J., Cermak, M.,
  La{\v{s}}tovi{\v{c}}ka, M.: {Enhancing Cybersecurity Skills by Creating
  Serious Games}. In: {Proceedings of the 23rd Annual ACM Conference on
  Innovation and Technology in Computer Science Education}. pp. 194--199.
  ITiCSE 2018, Association for Computing Machinery, New York, NY, USA (2018)

\bibitem{205231}
Taylor, C., Arias, P., Klopchic, J., Matarazzo, C., Dube, E.: {CTF:
  State-of-the-Art and Building the Next Generation}. In: 2017 {USENIX}
  Workshop on Advances in Security Education ({ASE} 17). {USENIX} Association,
  Vancouver, BC (Aug 2017)

\bibitem{7943509}
{Urias}, V.E., {Van Leeuwen}, B., {Stout}, W.M.S., {Lin}, H.W.: {Dynamic
  cybersecurity training environments for an evolving cyber workforce}. In:
  {2017 IEEE International Symposium on Technologies for Homeland Security
  (HST)}. pp.~1--6 (2017)

\bibitem{DoDCTT}
{US Department of Defense}: {The Department of Defense Cyber Table Top
  Guidebook} (2018), \url{https://www.dau.edu/cop/test/DAU Sponsored
  Documents/The DoD Cyber Table Top Guidebook v1.pdf}, (Accessed on August 28,
  2020)

\bibitem{brno}
Val{\'u}{\u{s}}ek, M.: {Classification of Network Attacks and Detection
  Methods}. Tech. rep., Masaryk University, Czech Republic (2016)

\bibitem{kypo}
Vykopal, J., O{\u{s}}lej{\u{s}}ek, R., Celeda, P., Vizv{\'a}ry, M.,
  Tovar{\u{n}}{\'a}k, D.: {KYPO Cyber Range: Design and Use Cases}. In:
  Proceedings of the 12th International Conference on Software Technologies -
  Volume 1: ICSOFT. pp. 310--321 (Jan 2017)

\bibitem{vykopal2017lessons}
Vykopal, J., Vizv{\'a}ry, M., O{\u{s}}lej{\u{s}}ek, R., Celeda, P.,
  Tovar{\u{n}}{\'a}k, D.: {Lessons Learned From Complex Hands-on Defence
  Exercises in a Cyber Range}. In: 2017 IEEE Frontiers in Education Conference
  (FIE). pp.~1--8. IEEE (2017)

\bibitem{Wright_blackhat_17}
Wright, A.C.: {Orange Is The New Purple}. Blackhat conference presentation
  (2017),
  \url{https://www.blackhat.com/docs/us-17/wednesday/us-17-Wright-Orange-Is-The-New-Purple-wp.pdf},
  (Accessed on August 28, 2020)

\bibitem{6113347}
{Wu}, Z., {Ou}, Y., {Liu}, Y.: {A Taxonomy of Network and Computer Attacks
  Based on Responses}. In: {2011 International Conference of Information
  Technology, Computer Engineering and Management Sciences}. vol.~1, pp. 26--29
  (2011)

\bibitem{YAMIN2020101636}
Yamin, M.M., Katt, B., Gkioulos, V.: {Cyber Ranges and Security Testbeds:
  Scenarios, Functions, Tools and Architecture}. Computers \& Security
  \textbf{88},  101636 (2020)

\bibitem{yurcik2001different}
Yurcik, W., Doss, D.: {Different Approaches in the Teaching of Information
  Systems Security}. In: {Proceedings of the Information Systems Education
  Conference}. pp. 32--33 (2001)

\end{thebibliography}
